\documentclass[reprint,aps,prmaterials,superscriptaddress]{revtex4-2}

\usepackage[utf8]{inputenc}

\usepackage[english]{babel}

\usepackage{color}

\usepackage{threeparttable}

\usepackage{amsfonts,amsmath,amsthm,amssymb}

\usepackage[normalem]{ulem}

\usepackage{graphicx}

\graphicspath{{img/}}
\DeclareGraphicsExtensions{.pdf,.png,.jpg}

\usepackage[colorlinks]{hyperref}

\begin{document}

\title{Nature of native atomic defects in ZrTe$_5$ and their impact on the low-energy electronic structure}

\author{B.~Salzmann}
\altaffiliation{Corresponding author.\\ bjoern.salzmann@unifr.ch}
\affiliation{D{\'e}partement de Physique and Fribourg Center for Nanomaterials, Universit{\'e} de Fribourg, CH-1700 Fribourg, Switzerland}

\author{A.~Pulkkinen}
\affiliation{D{\'e}partement de Physique and Fribourg Center for Nanomaterials, Universit{\'e} de Fribourg, CH-1700 Fribourg, Switzerland}
\affiliation{School of Engineering Science, LUT University, FI-53850, Lappeenranta, Finland}

\author{B.~Hildebrand}
\affiliation{D{\'e}partement de Physique and Fribourg Center for Nanomaterials, Universit{\'e} de Fribourg, CH-1700 Fribourg, Switzerland}

\author{T.~Jaouen}
\affiliation{D{\'e}partement de Physique and Fribourg Center for Nanomaterials, Universit{\'e} de Fribourg, CH-1700 Fribourg, Switzerland}
\affiliation{Universit{\'e} Rennes, CNRS, Institut de Physique de Rennes - UMR 6251, F-35000 Rennes, France}

\author{S.~N.~Zhang}
\affiliation{Institute of Physics, {\'E}cole Polytechnique F\'{e}d\'{e}rale de Lausanne (EPFL), CH-1015 Lausanne, Switzerland}
\affiliation{National Centre for Computational Design and Discovery of Novel Materials MARVEL, {\'E}cole Polytechnique F\'{e}d\'{e}rale de Lausanne (EPFL), CH-1015 Lausanne, Switzerland}

\author{E.~Martino}
\affiliation{Institute of Physics, {\'E}cole Polytechnique F\'{e}d\'{e}rale de Lausanne (EPFL), CH-1015 Lausanne, Switzerland}

\author{Q.~Li}
\affiliation{Condensed Matter Physics and Materials Science Department, Brookhaven National Laboratory, Upton, New York 11973, USA}

\author{G.~Gu}
\affiliation{Condensed Matter Physics and Materials Science Department, Brookhaven National Laboratory, Upton, New York 11973, USA}

\author{H.~Berger}
\affiliation{Institut de Physique des Nanostructures, {\'E}cole Polytechnique F{\'e}d{\'e}rale de Lausanne (EPFL), CH-1015 Lausanne, Switzerland}

\author{O.~V.~Yazyev}
\affiliation{Institute of Physics, {\'E}cole Polytechnique F\'{e}d\'{e}rale de Lausanne (EPFL), CH-1015 Lausanne, Switzerland}
\affiliation{National Centre for Computational Design and Discovery of Novel Materials MARVEL, {\'E}cole Polytechnique F\'{e}d\'{e}rale de Lausanne (EPFL), CH-1015 Lausanne, Switzerland}

\author{A.~Akrap}
\affiliation{D{\'e}partement de Physique and Fribourg Center for Nanomaterials, Universit{\'e} de Fribourg, CH-1700 Fribourg, Switzerland}

\author{C.~Monney}
\affiliation{D{\'e}partement de Physique and Fribourg Center for Nanomaterials, Universit{\'e} de Fribourg, CH-1700 Fribourg, Switzerland}

\date{\today}

\begin{abstract}
Over the past decades, investigations of the anomalous low-energy electronic properties of ZrTe$_5$ have reached a wide array of conclusions. An open question is the growth method's impact on the stoichiometry of ZrTe$_5$ samples, especially given the very small density of states near its chemical potential.
Here we report on high resolution scanning tunneling microscopy and spectroscopy measurements performed on samples grown via different methods.
Using density functional theory calculations, we identify the most prevalent types of atomic defects on the surface of ZrTe$_5$, namely Te vacancies and intercalated Zr atoms. Finally, we precisely quantify their density and outline their role as ionized defects in the anomalous resistivity of this material. 
\end{abstract}

\maketitle

\textit{Introduction.} 
Zirconium Pentatelluride (ZrTe$_5$) is a material that first attracted attention for its anomalous transport properties at temperatures around 150~K, namely a peak in resistivity and a simultaneous change of the sign of the Hall coefficient \cite{wieting1980giant, Izumi1982}.
Surprisingly, its chemical potential shifts from the valence band at room temperature to the conduction band at low temperature, suggesting an unknown intrinsic source of charge \cite{Zhang2017}. Although the material has been investigated for decades \cite{DiSalvo1981, Okada1982, Izumi1987, Rubinstein1999}, no consensus has emerged on either the relevant details of its electronic structure, nor on its exact topological classification. Various recent studies using different measurement and simulation techniques have come to a wide range of conclusions concerning the nature of ZrTe$_5$ and the resistivity anomaly.

In recent studies, ZrTe$_5$ has been classified as a semiconductor \cite{Shahi2018} with a potential transition to a semimetal \cite{mcilroy2004,Xu2018}, a three-dimensional (3D) Dirac semimetal \cite{Li2016, Chen2015}, a Weyl-semimetal \cite{liang2018}, and additionally as both a weak \cite{Zhang2017} and a strong \cite{Manzoni2016} topological insulator.
Experiments contradicting both the classification as a 3D Dirac semimetal \cite{martino2019} and as a strong topological insulator \cite{Wu2016, Xiang2016} have also been published.
One source of confusion appears to be variations between samples due to different growth techniques, namely Chemical Vapor Transport (CVT) and Flux methods. This can be addressed on a microscopic level with scanning tunneling microscopy (STM), as demonstrated recently for the transition metal dichalcogenide TiSe$_2$ \cite{Hildebrand2014}.
However, previous STM and scanning tunneling spectroscopy (STS) studies on ZrTe$_5$ have focused on investigating the presence of surface states as a signature of topological properties \cite{Xiang2016, Wu2016,Manzoni2016}, showing small and defect-free topographic images, and are relatively sparse in general due to difficulties in obtaining cleaved surfaces well suited for STM measurements. Although changes of the crystal lattice parameters induced by native defects have already been proposed \cite{Zhang2017, Manzoni2016}, a direct comparison between samples grown with these two methods has so far not been explored with the full array of available techniques \cite{Shahi2018}. While charge localization at defect sites has been proposed as an origin of the seeming violation of charge conservation with changing temperature, no study of the specific defects present in ZrTe$_5$ samples has been published thus far \cite{Zhang2017}.

Here, we present comparative STM and STS measurements on samples grown with both Flux and CVT methods. These allow insight into both structural and spectroscopic information up to atomic resolution.
With the help of simulated STM images calculated within the density functional theory (DFT), we identify two types of repeating point-defects.
Namely, Te vacancies and intercalated Zr atoms, the latter one being unique to CVT samples.
We further describe a long-range chemical potential fluctuation in the electronic structure of the CVT samples that we identify as charge puddles.
Finally, we propose that the identified native surface defects could explain the anomalous chemical potential of ZrTe$_5$, since they potentially represent an important source of intrinsic doping.

\begin{figure}[htbp]
\includegraphics[width=\linewidth]{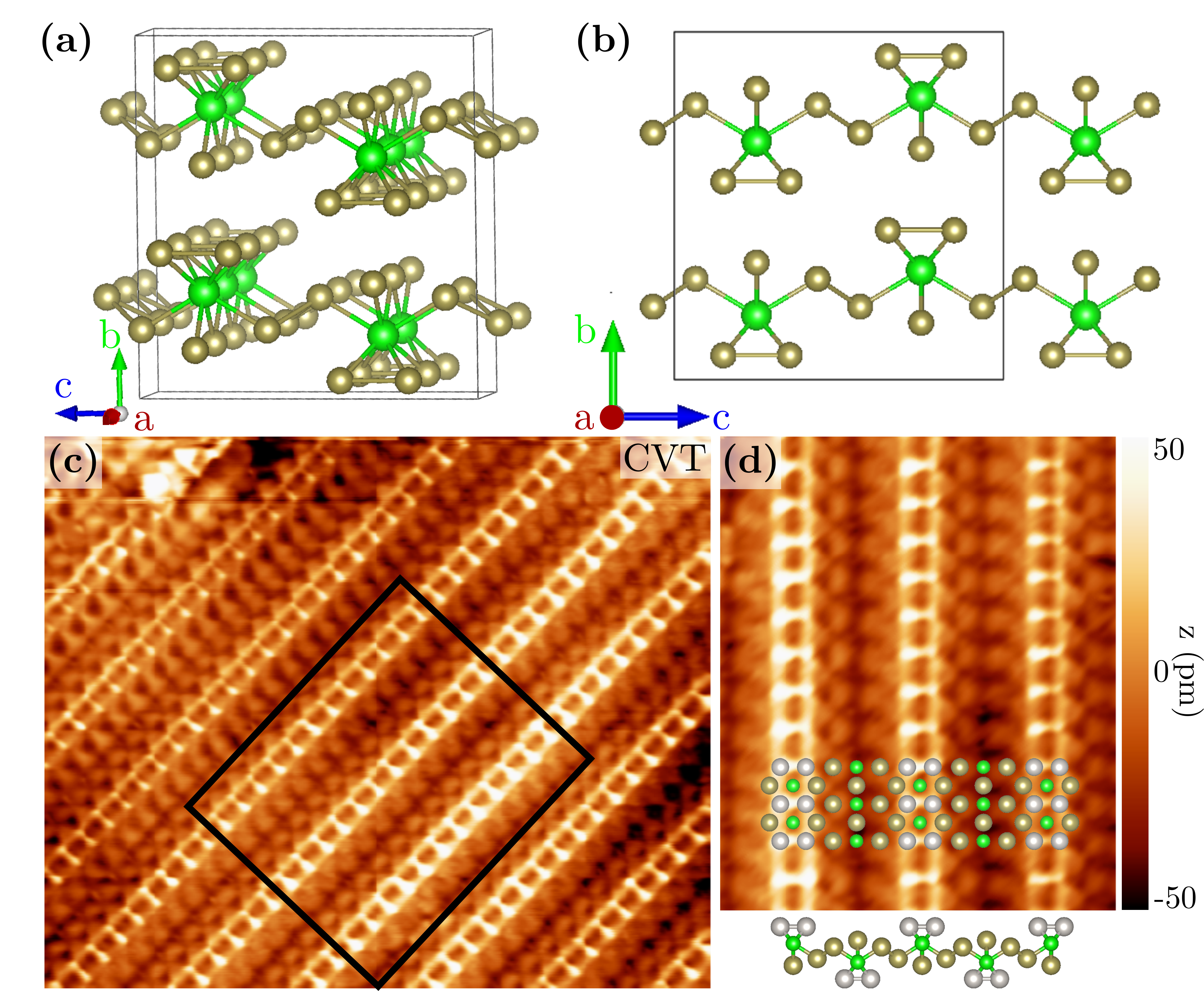} 
\caption{(a), (b) Crystal structure of ZrTe$_5$ as seen from different directions. Te and Zr atoms are shown in gold and green respectively. (c) 8.3$\times$10~nm$^2$ STM image of the (010) surface of ZrTe$_5$ showing atomic resolution using a tunnelling current of 0.3~nA. (d) Enlarged and rotated region indicated in (c). The overlay shows the locations of atoms close to the surface. Below the image a single layer of ZrTe$_5$ as seen along the chain direction is shown to visualize the height profile of the surface. The pairs of Te atoms aligned perpendicular to the Zr chains are shown in silver both in the overlay and the profile for visual clarity.}
\label{Fig:Cryst_Struct}
\end{figure}

\textit{Methods.} 
Flux growth of ZrTe$_5$ occurs within a vacuum sealed ampoule, containing the elementary constituents of ZrTe$_5$, with a Zr:Te ratio much smaller than 1:5 \cite{Li2016}.
The Te melt begins to dissolve the solid Zr, acting as both the Flux as well as part of the desired product \cite{Canfield1992}.
CVT growth involves an additional vapor component, Iodine, to serve as a transport agent for the constituents.
The base components are placed in the ampoule in a ratio that is close or equal to the stoichiometric ratio of the desired product \cite{LevyGrowth,Shahi2018, Wu2016}.
The STM and STS measurements were performed using an Omicron low temperature STM system in fixed current mode, with a bias applied to the sample.
All measurements were made at a temperature of 4.5~K, with a fixed current of 0.2~nA unless otherwise noted.
Samples were cleaved using scotch tape under vacuum (10$^{-8}$~mbar), before being transferred into ultra-high vacuum in the STM (\textless 10$^{-11}$~mbar) and cooled down to the measurement temperature.

DFT calculations were performed using the VASP code~\cite{kresse1993,kresse1994,kresse1996a,kresse1996b,kresse1999} within the projector augmented wave method~\cite{blochl1994}. 
The exchange and correlation effects were treated within the generalized gradient approximation using the PBE functional~\cite{perdew1996}. 
The ZrTe$_5$ surface was modeled with a 2-layer supercell with 6$\times$2 periodicity in the surface plane, corresponding to simulation cell dimensions 24.28~\r{A}$\times$27.67~\r{A}.
The atoms in the bottom ZrTe$_5$ layer were fixed to their bulk positions, and the top layer atoms were allowed to relax until the forces were converged to 0.02~eV/\r{A}. In the structural relaxations, the Brillouin zone was sampled using a Monkhorst-Pack grid of 3$\times$3$\times$1 k-points, and the kinetic energy cutoff was set to 400~eV. The simulated STM images were generated using the Tersoff-Hamann approach~\cite{tersoff1983}.

\textit{Results.} 
Figure~\ref{Fig:Cryst_Struct} shows a comparison between a small size atomically-resolved STM image and the atomic structure of ZrTe$_5$, made of layers stacked along the \textit{c} direction, see Fig. \ref{Fig:Cryst_Struct}(a) and (b).
Each layer is comprised of chains of ZrTe$_3$ running along the \textit{a} direction and connected by two Te atoms.
Due to the relatively weak interlayer bonding, the material cleaves easily parallel to the \textit{a-c} plane, exposing the (010) surface.
Figures \ref{Fig:Cryst_Struct}(c) and (d) show high-resolution STM images of the occupied states taken on a CVT sample at a bias voltage $U_{\text{bias}}=-0.3$~V with atomic resolution.
The comparison with the structural model in (d) highlights a very good correspondence between the expected position of the surface atoms and the image, allowing to identify the brightest parts of the STM image with the pairs of Te atoms atop the Zr chains.

\begin{figure}[htbp]
\centering
\includegraphics[width=\linewidth]{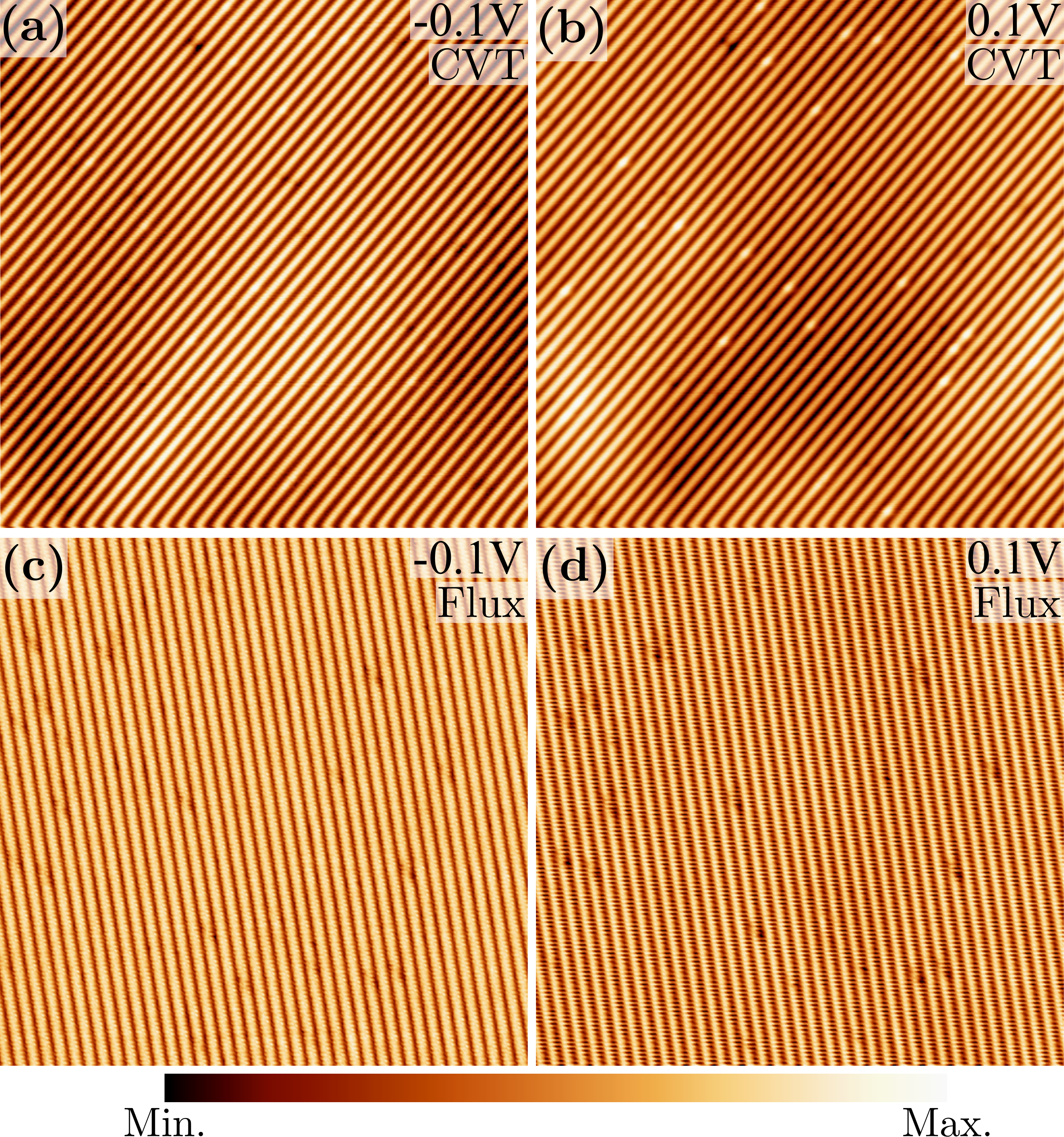}
\caption{Large scale topographic STM images taken on samples of both types. (a), (b) STM images taken at the same location on a CVT ZrTe$_5$ sample at different bias voltages. Both images are 50$\times$50~nm$^2$ in size. (c), (d) Two STM images taken at the same location on a Flux ZrTe$_5$ sample at the same bias voltages as (a) and (b). Both images are 50$\times$50~nm$^2$ in size. See Supplementary Materials for STM images over a full set of bias voltages.}
\label{Fig:STM_Comp}
\end{figure}

Figure~\ref{Fig:STM_Comp} shows large-scale STM images taken on both CVT and Flux samples with $U_{\text{bias}}=\pm 0.1$~V. We observe two main deviations from a perfectly ordered sample. First, there is a long-range variation in the intensity on the surface of CVT samples (see Fig. \ref{Fig:STM_Comp}(a) and (b)), with different regions tens of nanometers in size appearing brighter or darker.
These variations are absent on Flux samples (see Fig. \ref{Fig:STM_Comp}(c) and (d)). Second, there are two types of small and sharp defects that appear frequently in the STM images. We associate them with different types of atomic defects that occur either in or just below the surface layer. While the dark defects are found in both samples, the bright ones are only present in CVT samples. 
Similar STM images as those of Fig. 2 have been obtained on different surfaces of the same samples, as well as from other samples from the same batch, confirming the reproducibility of these results.

\begin{figure}[htbp]
\centering
\includegraphics[width=\linewidth]{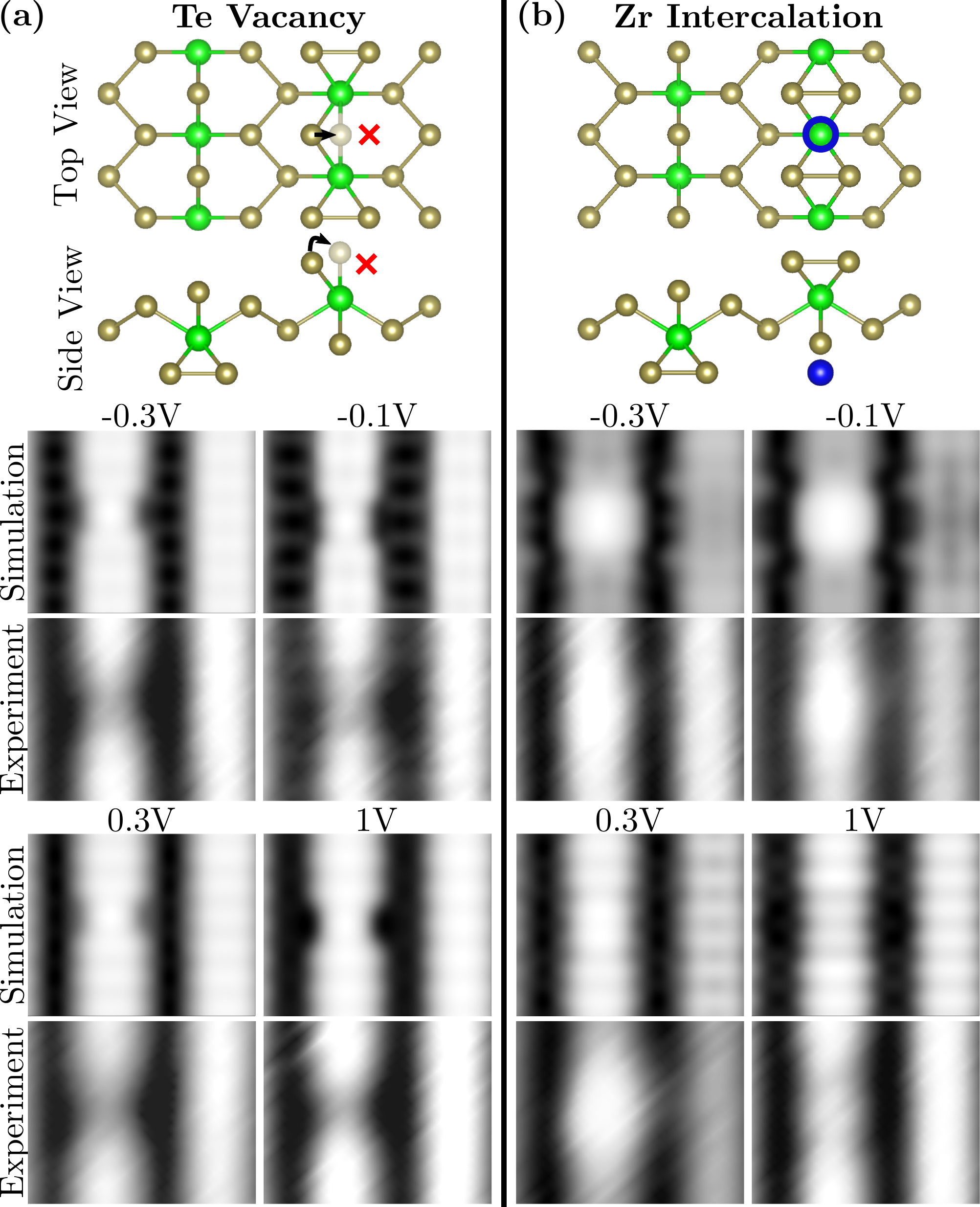}
\caption{Comparison between theoretical calculations and experimental images for the bright and dark surface defects. From top to bottom: Model of surface atomic structure in the vicinity of the defect with the location of the defect marked, model of defect as seen along the chain direction, simulated STM images of the defect at four different bias voltages and measured image of the defect at those voltages. (a) Missing Te atom from the top of a chain. The relaxation of the remaining Te atom toward the center of the chain is shown schematically. (b) A Zr atom intercalated below the top layer. The intercalated atom is shown in blue to distinguish it from the Zr atoms of the regular lattice.}
\label{Fig:Defects_HQ}
\end{figure}

To identify the nature of these defects, we show in Fig.~\ref{Fig:Defects_HQ}(a) a comparison between small-scale STM images and simulated STM images for different types of defects at $U_{\text{bias}}=-0.3$~V, $-0.1$~V,$0.3$~V and $1.0$~V \cite{SM}. The first type of defect, shown in Fig.~\ref{Fig:Defects_HQ}(a), appears as a reduction in intensity of the Zr chain at all bias voltages. We identify it as a Te vacancy in the top surface chains. The simulated images exhibit a strong agreement with the experimental data. They were obtained by removing one of the surface Te atoms from the top ZrTe$_3$ chain, as indicated in the top structural model (see Fig.~\ref{Fig:Defects_HQ}(a)), and subsequently allowing the structure to relax. The relaxation is an important step, as it allows the remaining side Te atom to relax toward the middle of the surface chain, an energetically more favorable position \cite{SM}. This results in a simulated defect with a more symmetric surface electronic density, in agreement with the experiment.
The second type of defect, shown in Fig.~\ref{Fig:Defects_HQ}(b), appears as a local increase in the intensity of the Zr chain for negative bias voltages and positive bias voltages up to around +0.4~V. At higher positive bias, the defect instead appears as a lowering of intensity at the same location. This specific behavior with bias voltage allows us to identify the defect as an intercalated Zr atom located directly below one of the Zr atoms in the chains.

We now turn back to the investigation of the long-range intensity variations observed at the surface of CVT samples, see Fig.~\ref{Fig:STM_Comp}(a) and (b). These intensity variations are clearly visible at $U_{\text{bias}}=+0.1$~V and -0.1~V, but are absent at higher voltages \cite{SM}. This indicates that they are not topographic in nature. In that case, they would appear at all voltages. Rather, they appear to be an electronic phenomenon confined to energies close to the chemical potential ($\mu$).
In the images taken at $U_{\text{bias}}=-0.1$~V and +0.1~V, the intensities of the long-range variations invert. Brighter regions at one voltage appear darker at the other voltage, and vice versa. This further supports the interpretation of these variations being purely electronic.

\begin{figure}[htbp]
\centering
\includegraphics[width=\linewidth]{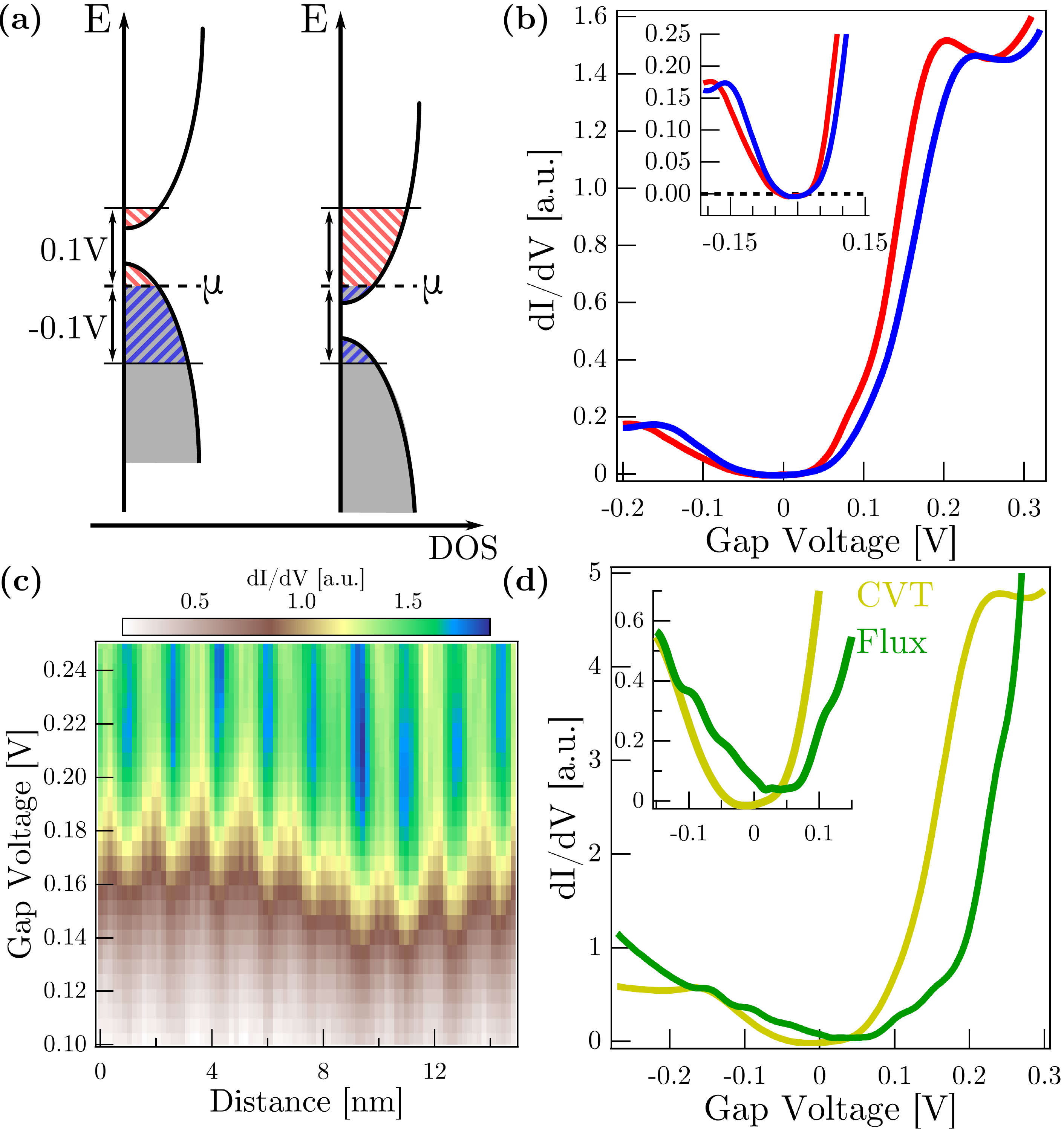}
\caption{(a) Diagram showing the proposed origin of the long-range intensity variation (see text). (b) Comparison between STS spectra averaged over two areas of 3.75$\times$3~nm$^2$ each, taken on a CVT sample showing opposite intensity variations. The spectra are shifted by 39~meV with respect to each other. (c) Map of STS spectra taken along a line across the transition between two areas with opposite intensity variation from the same STS map used in (b). The periodic pattern also visible in the spectra is due to the differences in the DOS on top of and between the Zr chains. (d) Comparison between STS spectra averaged over large areas on both CVT and Flux samples. The full STS maps used to obtain these data are shown in the Supplementary Materials.}
\label{Fig:STS_Spectra}
\end{figure}

These intensity variations can be understood considering the ZrTe$_5$ density of states (DOS) shown schematically in Fig.~\ref{Fig:STS_Spectra}(a). The total DOS exhibits a minimum close to $\mu$, and increases for both higher and lower energies within the relevant energy interval of $\pm$0.1~V. Therefore, if the chemical potential shifts locally with respect to that minimum, 
STM measurements at positive and negative bias voltages will receive unequal contributions from the sample's integrated DOS. Whether the positive or negative voltage shows the region as brighter is then determined by the direction in which the chemical potential is shifted with respect to the local minimum in the DOS. 

To support this scenario, we have performed STS measurements on CVT ZrTe$_5$, which indeed display local variations in the DOS, as exemplified in Fig.~\ref{Fig:STS_Spectra}(b) and (c).
Figure \ref{Fig:STS_Spectra}(b) shows the spatially averaged dI/dV curves as a function of $U_{\text{bias}}$ within two different regions of the same STS map \cite{SM}. Evidently, the DOS within the two regions are shifted with respect to each other, while otherwise showing the same features. Further, the overall shape of the DOS agrees with that seen in previous experiments \cite{Xiang2016}. Figure \ref{Fig:STS_Spectra}(c) shows a color map of several STS spectra taken along a line across the same STS map used for the average spectra in (b). It reveals a shift in the location of the plateau in the unoccupied states at $U_{\text{bias}}=0.2$~V over a distance of 6~nm.

\textit{Discussion.} 
The bias voltage behavior of the long-range intensity variation and their profile in STS is very reminiscent of similar observations made in graphene \cite{Samaddar2016,Beidenkopf2011} and topological insulators like BiSbTeSe$_2$ \cite{Knispel2017}. In semimetals with a small density of states near the chemical potential or in small gap semiconductors, the electric potential of charge impurities$\--$like ionized donors or acceptors in topological insulators$\--$is weakly screened and gives rise to the so-called Coulomb disorder, i.e. a strong local band bending with fluctuating potential \cite{Skinner2012}. This local band bending can bring the valence or conduction band of a compensated semimetal or semiconductor across the chemical potential, leading to a local charge accumulation, called charge puddles. The spatial size of these charge puddles typically depends on the gap size, dielectric constant and density of charge impurities. In the case of ZrTe$_5$, these charge puddles have typically a diameter of about 15~nm at 4.5~K in CVT samples. Interestingly, there are no charge puddles in our Flux samples. This can be explained by looking at their average STS spectra (away from charge puddles) in Fig.~\ref{Fig:STS_Spectra}(d), showing a comparison between the spatially averaged DOS of a CVT and a Flux sample. At 4.5~K, while the chemical potential lies in the band gap at the surface of CVT samples, it is shifted inside the valence band at the surface of the Flux samples. Therefore, we expect more charge carriers at the surface of Flux samples, leading to an increased screening that suppresses the occurrence of charge puddles.
While at first it seems compelling to associate the appearance of these charge puddles on the CVT samples with the influence of the bright atomic defects, no clear correlation between their location and the charge puddles can be seen in our STM images (see Fig.~\ref{Fig:STM_Comp}).

We now turn to the quantitative  discussion of atomic defects. From a comparison between STM images and DFT calculations, Te vacancies (dark defects) were identified in both Flux and CVT samples, while Zr intercalated atoms (bright defects) were observed only in CVT samples (see Fig.~\ref{Fig:STM_Comp}).
Te vacancies appear at a rate of (4.5$\pm$0.4)$\:\cdot\:$10$^{-3}$ and (5.2$\pm$0.4)$\:\cdot\:$10$^{-3}$ per unit cell when averaged over large imaged areas (more than 10'000~nm$^2$ each) for CVT samples and Flux samples, respectively.
The average frequency of the intercalated Zr atoms on CVT samples is (2.1$\pm$0.3)$\:\cdot\:$10$^{-3}$ per unit cell. Altogether, the Zr to Te ratio \textit{near the surface} for both types of samples amounts to 1:4.98$\pm$0.01 and 1:4.995$\pm$0.001 in CVT and Flux samples respectively. We stress here that the tiny difference between these two ratios is significant, and is essentially coming from the presence of intercalated Zr atoms in CVT samples only.
It is particularly interesting to relate our results to the work of Shahi \textit{et al.} \cite{Shahi2018}. They notably studied both types of samples with bulk sensitive methods like x-ray diffraction and energy dispersive x-ray spectroscopy, finding a deficiency of Te in both CVT and Flux samples with a ratio of about 1:4.60 $\pm 0.20$ and 1:4.98 $\pm 0.17$, respectively. This is consistent with our conclusions, since this bulk study also evidences a better stoichiometry in Flux samples with our study providing direct evidence of the defects causing the deviation from a perfect stoichiometry.

What is the influence of the atomic defects on the electronic structure of ZrTe$_5$? As discussed previously, our STS measurements (see Fig.~\ref{Fig:STS_Spectra}(d)) show a chemical potential difference between CVT and Flux grown samples. This difference corresponds to additional electron doping in CVT samples, consistent with the presence of intercalated Zr atoms (partially) giving their $4d$ and $5s$ electrons, in agreement with our calculations of the defect-related doping of the two samples above. 
It is very likely that the atomic defect states common to both types of samples (Te vacancies) play an important role for transport properties, in particular for the anomalous chemical potential shift of ZrTe$_5$. Indeed, states close to the chemical potential have mostly Te character \cite{Manzoni2016}, therefore such defect states could be easily activated even at low temperatures, leading to the change in dominant charge carrier type \cite{Zhang2017}. Finally, the electron doping due to Zr intercalated atoms accounts for the shift in temperature of the resistivity peak in CVT samples, in comparison to Flux samples and also the appearance of charge puddles in CVT samples due to a shift of their chemical potential in the gap. \cite{SM}.

\textit{Conclusion.} 
In this letter, we have identified native defects for both ZrTe$_5$ samples grown by CVT and Flux methods and demonstrated their relevance to the bulk properties of the material.
These include not only point-defects, but also charge puddles appearing in CVT grown ZrTe$_5$.
The above results clearly show the importance of taking into account differences between samples produced by different growth methods.
They also invite further investigation on the influence of growth method on the quality of the produced samples, and from there on the electronic and topological character of ZrTe$_5$.

\textit{Acknowledgments.} 
This project was supported by the Swiss National Science Foundation (SNSF) Grant No. P00P2\_170597. A.P. acknowledges the Osk. Huttunen Foundation for financial support, and CSC – IT Center for Science, Finland, for computational resources. B.S. acknowledges support from MaNEP through its Summer internship grants. S.N.Z. and O.V.Y. acknowledge support by NCCR Marvel. A.A. acknowledges funding by the SNSF, Grant No. P00P2$\_$170544. Work at BNL is supported by US DOE, DE-SC0012704. We are very grateful to P.~Aebi for fruitful discussions and for sharing with us his STM setup. We acknowledge M.~Rumo and G.~Kremer for help in preparing samples. Skillful technical assistance was provided by F.~Bourqui, B.~Hediger and O.~Raetzo.

\onecolumngrid
\newpage
\begin{center}
{\large\textbf{\boldmath
Supplemental Material\\[0.5em] {\small for}\\[0.5em] Nature of native atomic defects in ZrTe$_5$ and their impact on the low-energy electronic structure}}\\[1.5em]

B.~Salzmann,$^1$ A.~Pulkkinen,$^{1, 2}$ B.~Hildebrand,$^1$ T.~Jaouen,$^{1, 3}$ S.~N.~Zhang,$^{4, 5}$\\
E.~Martino,$^4$, Q.~Li$^6$, G.~Gu$^6$, H.~Berger$^7$, O.~V.~Yazyev$^{4, 5}$, A.~Akrap$^1$ and C.~Monney$^1$\\[0.5em]

\textit{\small
$^1$D{\'e}partement de Physique and Fribourg Center for Nanomaterials,\\ Universit{\'e} de Fribourg, CH-1700 Fribourg, Switzerland\\
$^2$School of Engineering Science, LUT University, FI-53850, Lappeenranta, Finland\\
$^3$Universit{\'e} Rennes, CNRS, Institut de Physique de Rennes - UMR 6251, F-35000 Rennes, France\\
$^4$Institute of Physics, {\'E}cole Polytechnique F\'{e}d\'{e}rale de\\
Lausanne (EPFL), CH-1015 Lausanne, Switzerland\\
$^5$National Centre for Computational Design and Discovery of Novel Materials MARVEL,\\
{\'E}cole Polytechnique F\'{e}d\'{e}rale de Lausanne (EPFL), CH-1015 Lausanne, Switzerland\\
$^6$Condensed Matter Physics and Materials Science Department,\\
Brookhaven National Laboratory, Upton, New York 11973, USA\\
$^7$Institut de Physique des Nanostructures,\\
{\'E}cole Polytechnique F{\'e}d{\'e}rale de Lausanne (EPFL), CH-1015 Lausanne, Switzerland
}

\vspace{1em}
\end{center}\setcounter{figure}{0}
\setcounter{equation}{0}
\renewcommand{\thefigure}{SM\arabic{figure}}
\renewcommand{\theequation}{SM\arabic{equation}}

{\center\section{Experimental Data}}

\begin{figure}[htb]
\centering
\includegraphics[width=\linewidth]{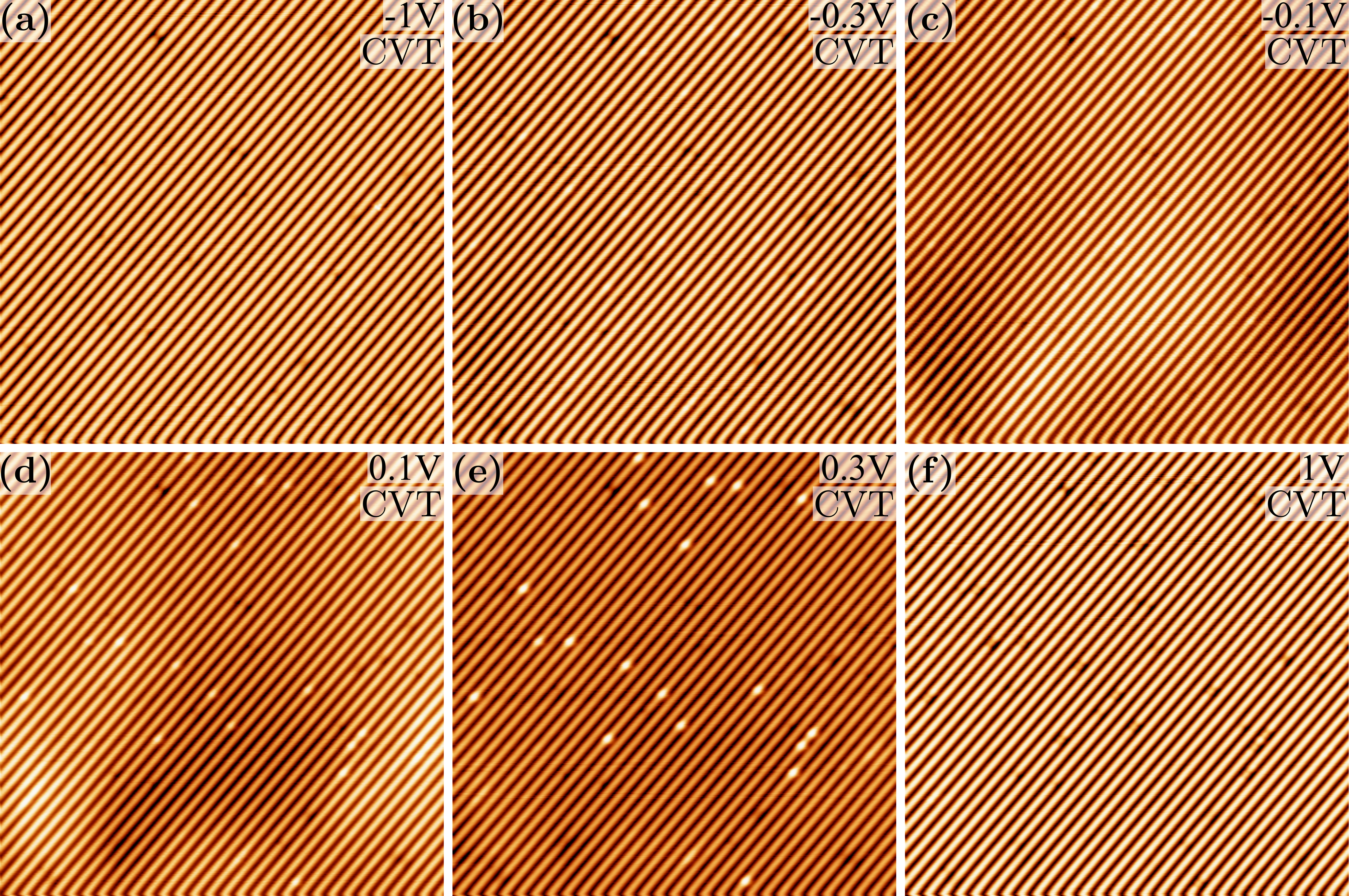}
\caption{Full voltage series measured on a CVT grown sample from -1 V to +1 V taken at the location of the STM images in Figure 2(a) and (b) in the main text. All images show a 50x50 nm$^2$ area imaged with a fixed tunneling current of 0.2 nA and at 4.5~K.}
\label{Fig:Add_STM_CVT}
\end{figure}

\begin{figure}[htb]
\centering
\includegraphics[width=\linewidth]{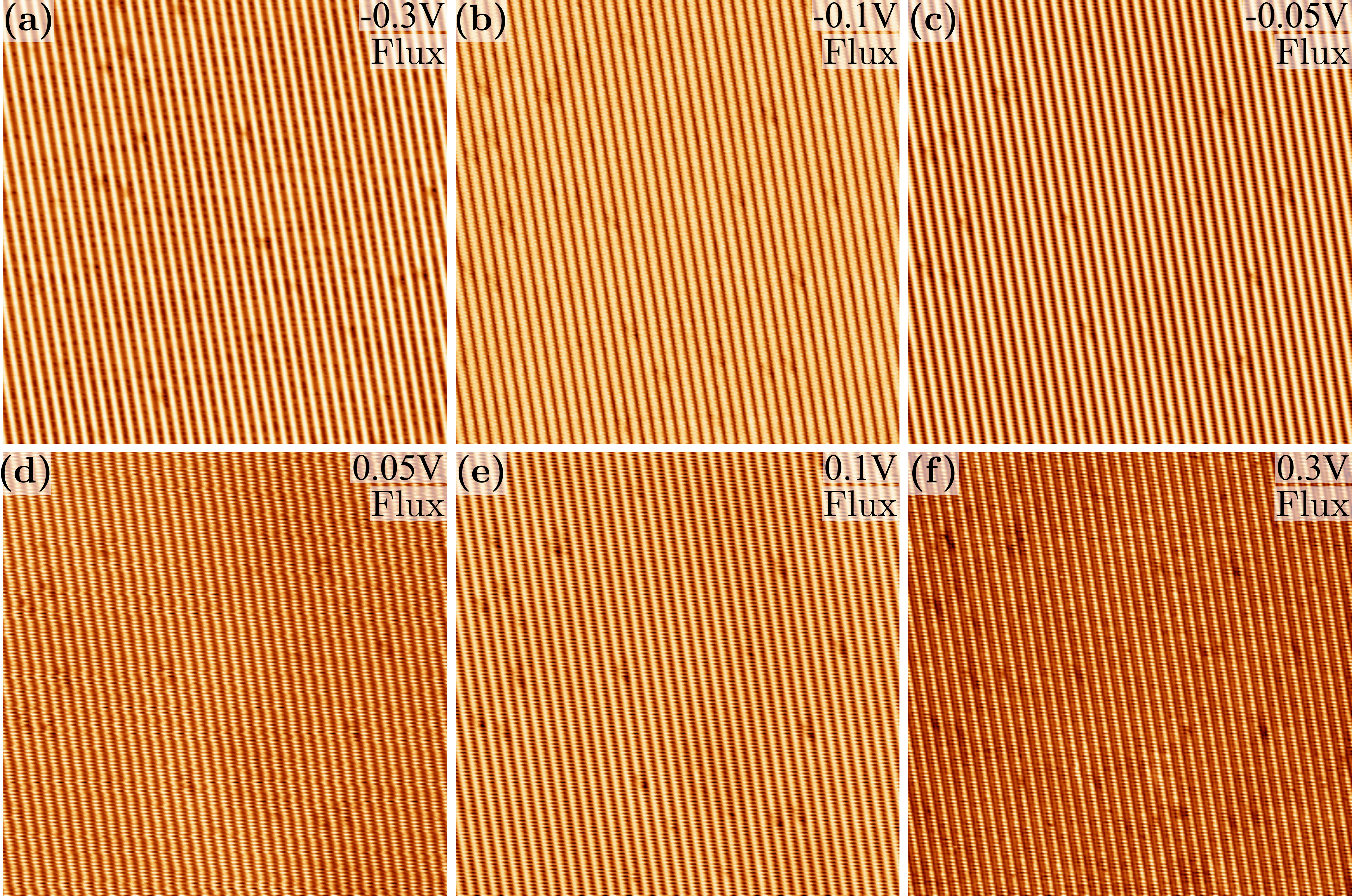}
\caption{Full voltage series measured on a Flux grown sample from -0.3 V to +0.3 V taken at the location of the STM images in Figure 2(c) and (d) in the main text. All images show a 50x50 nm$^2$ area imaged with a fixed tunneling current of 0.2 nA and at 4.5~K.}
\label{Fig:Add_STM_Flux}
\end{figure}

Figure \ref{Fig:Add_STM_CVT} shows a series of STM images at bias voltages from -1 V to +1 V taken at the same location on a CVT grown sample of ZrTe$_5$.
As mentioned in the main text, the long range intensity variations visible at $\pm$0.1 V do not appear at higher voltages.
Additionally, one recognizes two different types of recurring atomic defects from their intensity variation as a function of bias voltage, as described in the main text.

Figure \ref{Fig:Add_STM_Flux} shows a series of STM images at bias voltages from -0.3 V to +0.3 V taken at the same location on a Flux grown sample.
Even at $\pm$0.05 V bias voltage, no long range intensity variation is visible.

\begin{figure}[htbp]
\centering
\includegraphics[width=0.7\linewidth]{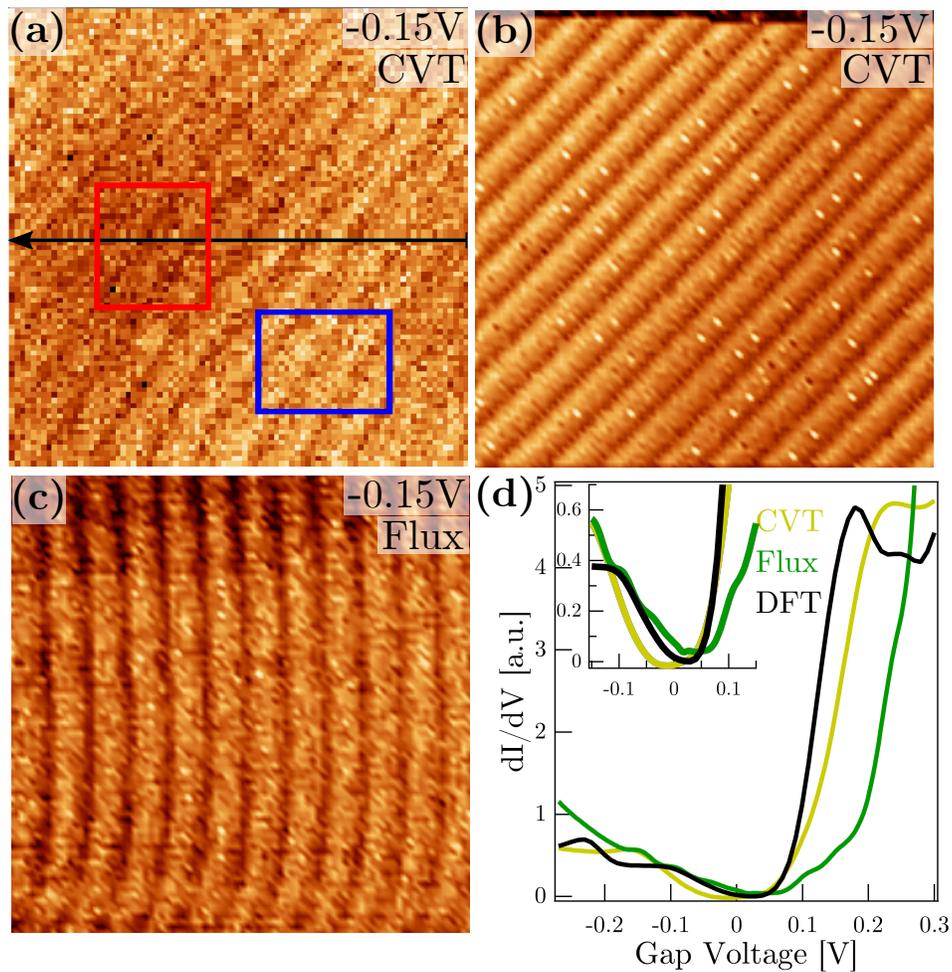}
\caption{(a) STS map showing a long range variation in intensity taken on a CVT grown sample. The map has a size of 15x15 nm$^2$ and was taken at a tunneling current of 0.2 nA.(b) STS map whose spectra were averaged to produce the CVT spectrum in figure 4(d) of the main text. The map has a size of 15x15 nm$^2$ and was taken at a tunneling current of 0.15 nA and at 4.5~K.(c) STS map taken on a Flux grown sample. The map has a size of 15x15 nm$^2$ and was taken at a tunneling current of 0.13 nA.(d) Comparison between the averaged STS spectra from a CVT and a Flux grown sample and the DOS from DFT calculations. The inset shows the energy range around the Fermi energy enlarged.}
\label{Fig:STS_Sources}
\end{figure}

Figure \ref{Fig:STS_Sources}(a) and (b) shows two STS maps taken on CVT grown samples.
Spectra from these maps were used in figure 4 in the main text.
The map in Figure \ref{Fig:STS_Sources} (a) shows a long range variation in intensity caused by a shifting of the DOS as described in the main text.
The locations averaged for the dI/dV spectra in figure 4(b) of the main text are marked with boxes in their respective colors.
The position of the line spectrum in figure 4(c) in the main text is shown in black.
Figure \ref{Fig:STS_Sources}(b) shows the STS map whose spectra were averaged into the dI/dV curve representing the CVT sample in figure 4(d) of the main text.
Figure \ref{Fig:STS_Sources}(c) shows an STS map taken on a Flux grown sample.
The locations of the spectra averaged for the dI/dV curve representing the Flux sample in figure 4(d) of the main text are indicated by the green box.
Figure \ref{Fig:STS_Sources}(d) contains a comparison between experimental dI/dV curves and calculated DOS from DFT calculations.
The general behaviour of the DOS is reproduced correctly and the location of the Fermi level in the calculations is well aligned with the experimental data.

\begin{figure}[htbp]
\centering
\includegraphics[width=0.6\linewidth]{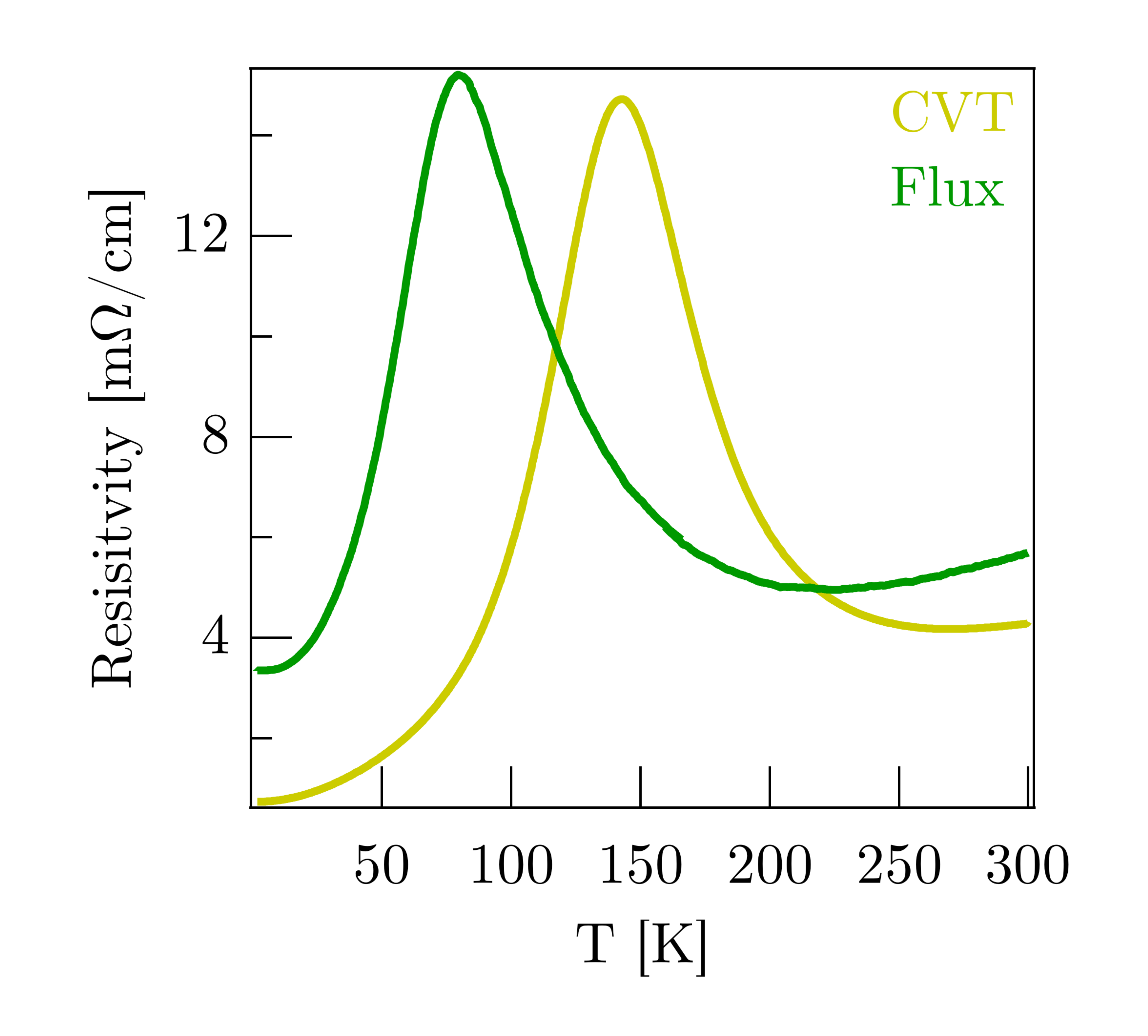}
\caption{Resistivity as a function of temperature for samples grown by the two methods, both showing a peak in resistivity at different temperatures.}
\label{Fig:Resistivity}
\end{figure}

Figure \ref{Fig:Resistivity} shows resistivity data for both sample types obtained on samples of the same batch as those used for the STM measurements.
The resistivity peak for the Flux grown sample appears at lower temperatures than that of the CVT grown sample.

{\center\section{DFT Calculations}}

\begin{figure}[htbp]
\centering
\includegraphics[width=\linewidth]{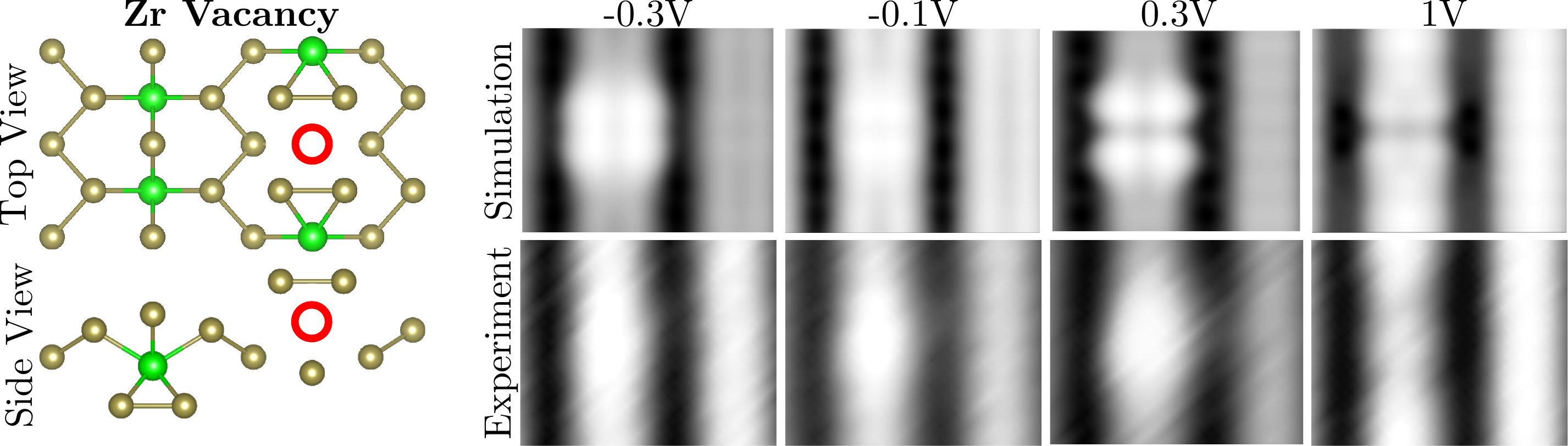}
\caption{Comparison between DFT calculation and experiment for a second proposed origin for the bright atomic defects seen on CVT grown samples. From left to right: structural model of the Zr vacancy defect, comparison between calculated and experimental images at different bias voltages.}
\label{Fig:ZrVacancy_DFT}
\end{figure}

In addition to the Te vacancy and Zr intercalated atom presented in the main text, we also investigated a Zr vacancy as a potential defect occurring in ZrTe$_5$ in the same manner as the other two defects, by comparison with DFT simulations.
Figure \ref{Fig:ZrVacancy_DFT} shows the comparison between these simulations and the experimental data at the same voltages as in the main text.
We find that the correspondence between calculation and experiment is worse for this defect than for the intercalated Zr atom.
Namely, the gap in the chain at the location of the defect seen in the simulations at 0.3~V and 1~V in figure \ref{Fig:ZrVacancy_DFT} does not appear in the measurements.
At the same time, the relative brightness of the defect at -0.3~V and -0.1~V seen in the experiment is not well reproduced by the calculations.
Most importantly, the calculations for the Zr vacancy do not reproduce the change from the defect being brighter than the surrounding chain to being darker when moving from 0.3~V to 1~V bias voltage.

\begin{figure}[htbp]
\centering
\includegraphics[width=\linewidth]{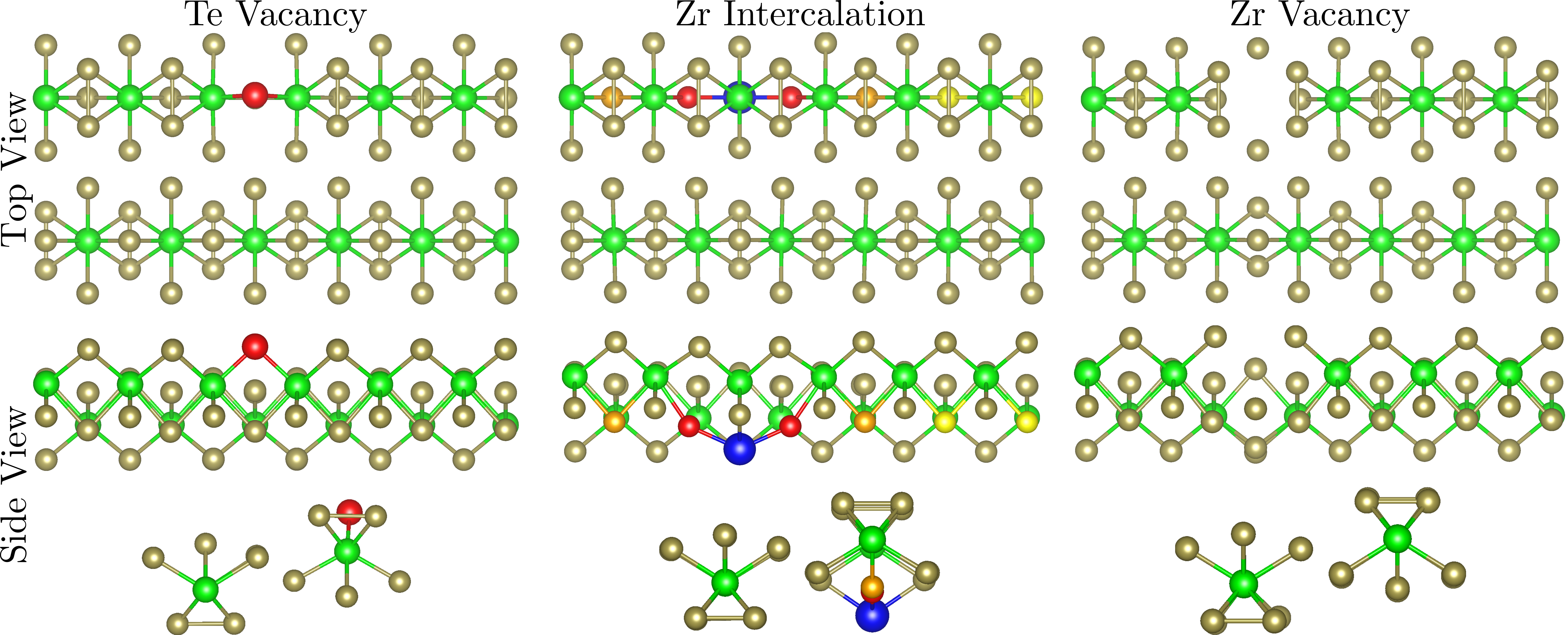}
\caption{Relaxed structures in the regions around all three studied defects as seen in, from top to bottom, \textit{b}, \textit{c} and \textit{a} directions. Atoms strongly displaced from their original positions are highlighted in different colors.}
\label{Fig:Relaxed_Structure}
\end{figure}

Before simulating the STM images of the different defects, the crystal structure was allowed to relax in response to the introduced defect.
Figure \ref{Fig:Relaxed_Structure} shows the resulting relaxed structures for each of the studied defects.
The way the structure relaxes gives insight into the range of the lattice effects of the defect.
In this context, the comparison between the Zr vacancy and the intercalated Zr again favours the Zr intercalation which shows a strong effect on the atomic position of Te atoms at longer distances away from the defect in the \textit{a} direction which are absent for the Zr vacancy.
This is best seen in the top view in figure \ref{Fig:Relaxed_Structure}.
When looking at the size of the defect visible in STM images, there is clearly a significant range of the defect along this direction.
The defect size in the \textit{a} direction at 0.3~V bias voltage where the defect appears largest is around 11~\r{A}, close to three unit cells in the \textit{a} direction (11.9~\r{A}), which agrees well with the lattice distortion in the simulation of the Zr intercalation whereas the effects of the Zr vacancy are more short ranged.
\end{document}